\mag=\magstephalf
\pageno=1
\input amstex
\documentstyle{amsppt}
\TagsOnRight

\pagewidth{16.5 truecm}
\pageheight{23.0 truecm}
\vcorrection{-1.0cm}
\hcorrection{-0.5cm}
\nologo
\NoRunningHeads
\NoBlackBoxes
\font\twobf=cmbx12

\define \res{\roman {res}}

\define \CC{{\Bbb C}}

\define \al{\roman {al}}

\define\hw{\widehat{\omega}}

\define \tvskip{\vskip 1.0 cm}
\define\ce#1{\lceil#1\rceil}
\define\dg#1{(d^{\circ}\geq#1)}
\define\Dg#1#2{(d^{\circ}(#1)\geq#2)}

\def\fp{\flushpar}

\define\s#1{\sigma_{#1}}
\define\tp#1{\negthinspace\left.\ ^t#1\right.}
\define\mrm#1{\text{\rm#1}}
\define\lr#1{^{\sssize\left(#1\right)}}

\redefine\qed{\hbox{\vrule height6pt width3pt depth0pt}}
\font\Large=cmr10 scaled \magstep5

{\centerline{\bf{Hyperelliptic Solutions of
Modified Korteweg-de Vries Equation of Genus $g$:}}}

{\centerline{\bf{Essentials of Miura Transformation}}}

\author
Shigeki MATSUTANI
\endauthor
\affil
8-21-1 Higashi-Linkan Sagamihara 228-0811 Japan
\endaffil \endtopmatter

\footnotetext{e-mail:RXB01142\@nifty.ne.jp}

\document

\centerline{\twobf Abstract }\tvskip

Explicit hyperelliptic solutions of the
 modified Korteweg-de Vries equations without
any ambiguous parameters were constructed
in terms only of the hyperelliptic $\al$-functions
over non-degenerated hyperelliptic curve $y^2 = f(x)$ of
arbitrary genus $g$. In the derivation,  any $\theta$-functions or
Baker-Akhiezer functions  were not essentially
used.
Then the Miura transformation naturally appears as
the connections between the hyperelliptic
$\wp$-functions and hyperelliptic $\al$-functions.

{\centerline{\bf{2000 MSC: 37K20, 35Q53, 14H45, 14H70 }}}

{\centerline{\bf{PACS: 02.30.Gp, 02.30.Dk, 02.30.Jr,
43.25.Rq
 }}}


\vskip 1 cm
{\centerline{\bf{\S 1. Introduction}}}
\vskip 0.5 cm

\vskip 1.0 cm

It might not be excess expressions that the discovery of
Miura transformation opened the modern soliton theory [GGKM].
Using it, the inverse scattering method was discovered
and developed [L].
In such developments of the theory,
there appeared several algebraic solutions of integrable
nonlinear differential equations in 1970's
([Kr], references in [BBEIM], [TD]).
These solutions were obtained by arguments on the behaviors
of algebraic or meromorphic functions
around infinite points of corresponding algebraic curves.
Their arguments are based on
a certain  existence theorems of holomorphic functions and
theorem of identity of analytic functions.
They played very important roles and contributed to the
developments of the soliton theories in 1980's. They
are especially connected to the abstract theories of
modern mathematics, which are, now, called
infinite dimensional analysis [SN, SS].
Thus, needless to say  that they have also important
meanings.

However their solutions are not, sometimes, enough to have
effects on some concrete computations of algebraic solutions;
{\it e.g.} the cases in which one wishes to determine coefficients
of higher degrees of Laurent
or Taylor expansions of the
solutions and in which one wishes to draw a graph of solutions without
computation of transcendental equations.
Recently theories of hyperelliptic functions which were
discovered in nineteenth century [Ba1-3, Kl1] has been
re-evaluated in various fields
[BEL1-3,CEEK, EE, EEK, EEL, EEP, G, Ma1-2, N, \^O1-2].
One of purpose of the re-evaluations seems to reconstruct
and to refine the modern soliton theory and algebraic geometry.

Because the theories of nineteenth century are very
concrete. First they chose a hyperelliptic curve. After
fixing the curve, they considered the meromorphic functions
over it. In nineteenth century, the study of the hyperelliptic
functions was very important as a generalization of
elliptic function, which was done by Weierstrass, Klein, Jacobi
and so on [D, Kl2].  As it should be very surprising,
even in the nineteenth century,
there appeared most of tools and objects in soliton
theories. In fact,
Baker found the Korteweg-de Vries hierarchy and
Kadomtsev-Petviashvili (KP) equations around 1898
in terms of bilinear operators, Paffians, symmetric functions
[Ba3, Ma2].

By changing the century now, we can look around both of them.
It is very interesting to compare the theory of
 soliton equations in the ending of last century with
that of hyperelliptic function in nineteenth century.
Both have merits and demerits.
When one wishes the concrete expressions, the theories of nineteenth
century are very convenient.
However if one wishes to deal with it in the framework of
infinite dimensional analysis, the older theories
might be useless.

This article is on the hyperelliptic solutions of
modified Korteweg-de Vries equation (MKdV)
related to hyperelliptic curves with higher genus
by employing the fashion in nineteenth century.
In the previous report [Ma3],
I constructed those of genus one and two associated with
loop solitons.
This article can be regarded as its sequel one to
arbitrary genus.
Only using the Weierstrass $\al$-function [W, Ba2 p.340], we will construct th
e
hyperelliptic solutions of the MKdV equation in theorem 3-2.
In the construction, we will not essentially use
any $\theta$ functions or Baker-Akhiezer functions
though the algebro-geometric solutions in terms of them
were already obtained by Gesztesy and Holden [GH1] recently.
Our solutions are explicitly constructed without
ambiguous parameters for arbitrary hyperelliptic curves and
our proof might be, I believe, simpler than that in terms of
Baker-Akhiezer functions. In the proof,
 we translate the differential operators in the Jacobian
into those in the curve and evaluate the related terms by using
residual computations and symmetries.
As the Weierstrass $\al$-function should be regarded as
a generalization of the Jacobi elliptic functions and
in a calculus  of Jacobi elliptic functions, we do not need
theta-functions except addition formula, our proof is
more natural.

In the construction of the hyperelliptic solutions of
the MKdV equation, we compare the hyperelliptic $\wp$
functions and the $\al$-functions.
As the hyperelliptic $\wp$ functions are connected with
the finite type solutions of the KdV equation [BEL1, 2, Ma2],
 we can investigate the Miura transformation between
finite type solutions of the KdV equations and MKdV equations.
In the Remark 3-3 and \S4, we show that
the Miura transformation for the finite type solutions
is also natural from several viewpoints.
 Here we should remark that
the comparison is essentially
the same as the Baker's arguments in [Ba3]; there
he found the KdV hierarchy and KP equation as
relations among the $\wp$ functions [Ba3, Ma2].
Further we note that even though they did not mention it,
Eilbeck, Enolskii and Kostov also found
similar relation between $\wp$ and $\al$ on their computations of
the vector nonlinear Schr\"odinger equation [EEK].

Due to [GGKM], any solutions of the KdV equation
can be regarded as a potential in a one-dimensional
Schr\"odinger equation whose spectrum is preserved
for the time development of the solutions. From the
spectral theory, the spectrum is  determined by a
characteristic relation over the complex number,
which is identified with a hyperelliptic curve for a certain
case. For the hyperelliptic curve case, the solutions
of the KdV equation are given as meromorphic functions
over the curve, which are called finite type solutions.
 Further  any characteristic relations
 can be approximated by the hyperelliptic
curves via the Weierstrass's approximation theorem.
It implies that the finite type solutions are dense in
the solutions space of the KdV equation. Thus
Our remarks in \S 3 and \S 4
 exhibit essentials of the Miura transformation.
Though Gesztesy and Holden [GH2] also considered the
Miura transformation from viewpoint of algebro-geometrical
computations of MKdV equation, our viewpoint on the remarks
differs from theirs. Thus our remarks give another aspect
of Miura transformation.

\vskip 0.5 cm

{\centerline{\bf{\S 2. Differentials of a Hyperelliptic Curve}}}

\vskip 0.5 cm

In this section, we will give the conventions which express
 the hyperelliptic functions in this article.
 As there is a good self-contained paper on
theories of hyperelliptic sigma functions [BEL2]
besides [Ba1,2,3],
we will give them without explanations and proofs.

We denote the set of complex number by $\Bbb C$ and
the set of integers by $\Bbb Z$.

\proclaim{\fp Convention 2-1}\it
We deal with a hyperelliptic curve $C_g$  of genus $g$
$(g>0)$ given by the affine equation,
$$ \split
   y^2 &= f(x) \\  &= \lambda_{2g+1} x^{2g+1} +
\lambda_{2g} x^{2g}+\cdots  +\lambda_2 x^2
+\lambda_1 x+\lambda_0  \\
     &=P(x)Q(x),\\
\endsplit  \tag 2-1
$$
where $\lambda_{2g+1}\equiv1$ and $\lambda_j$'s are
complex numbers. We use the expressions
$$
\split
        f(x) &= (x-b_1)(x-b_2)\cdots(x-b_{2g})(x-b_{2g+1}),\\
        Q(x) &= (x-c_1)(x-c_2)\cdots(x-c_g)(x-c),\\
        P(x) &= (x-a_1)(x-a_2)\cdots(x-a_g), \\
\endsplit \tag 2-2
$$
where $a_j$'s, $b_j$'s, $c_j$'s  and $c$ are  complex values
and $b_{2g+2}=\infty$.
\endproclaim

\proclaim{Proposition 2-2}

There exist
 several symmetries which express the same curve $C_g$.

\roster

\item translational symmetry: For $\alpha_0 \in\CC$,
$(x,y) \to (x+\alpha_0,y)$,
with $b_j \to b_j + \alpha_0$.

\item dilatation symmetry: For $\alpha_1 \in\CC$,
$(x,y) \to (\alpha_0x,\alpha_0^{2g+1}y)$,
with $b_j \to \alpha_0 b_j$.

\item inversion symmetry:  For fixing $b_i$,
$(x,y) \to (1/(x-b_i),y \prod_{j\neq i} \sqrt{b_i-b_j}/(x-b_i)^{(2g+1)/2})$
with
$b_j \to 1/(b_j-b_i)$.

\endroster
\endproclaim

\proclaim{\fp Definition 2-3 [Ba1 , Ba2 BEL2, \^O1]}\it

 \roster
For a point $(x_i, y_i)\in C_g$, we define the following
quantities.

\item  Let us denote the homology of a hyperelliptic
curve $C_g $ by
$$
\roman{H}_1(C_g, \Bbb Z)
  =\bigoplus_{j=1}^g\Bbb Z\alpha_{j}
   \oplus\bigoplus_{j=1}^g\Bbb Z\beta_{j},
 \tag 2-3
$$
where these intersections are given as
$[\alpha_i, \alpha_j]=0$, $[\beta_i, \beta_j]=0$ and
$[\alpha_i, \beta_j]=\delta_{i,j}$.

\item The unnormalized differentials of the first kind are
defined by,
$$   d u^{(i)}_1 := \frac{ d x_i}{2y_i}, \quad
      d u^{(i)}_2 :=  \frac{x_i d x_i}{2y_i}, \quad \cdots, \quad
     d u^{(i)}_g :=\frac{x_i^{g-1} d x_i}{2 y_i}.
      \tag 2-4
$$

\item The unnormalized complete hyperelliptic integral
of the first kind are defined by,
$$    \pmb{\omega}':=\left[\left(
     \int_{\alpha_{j}}d u^{(a)}_{i}\right)_{ij}\right],
\quad
      \pmb{\omega}'':=\left[\left(
       \int_{\beta_{j}}d u^{(a)}_{i}\right)_{ij}\right],
 \quad
    \pmb{\omega}:=\left[\matrix \pmb{\omega}' \\ \pmb{\omega}''
     \endmatrix\right].
  \tag 2-5
$$

\item The normalized complete hyperelliptic
integral of the first kind are given by,
$$    \ ^t\left[\matrix \hw_{1}  \cdots & \hw_{g}
        \endmatrix\right]
       :={\pmb{\omega}'}^{-1}  \ ^t\left[\matrix
          d u^{(i)}_{1} & \cdots
   d u^{(i)}_{g}\endmatrix\right] ,\quad
   \pmb \tau:={\pmb{\omega}'}^{-1}\pmb{\omega}'',
   \quad
    \hat{\pmb{\omega}}:=\left[\matrix 1_g \\ \pmb \tau
     \endmatrix\right].
       \tag 2-6
$$

\item The unnormalized differentials of the second kind are
 defined by,
$$   d \tilde u^{(i)}_1 := \frac{x_i^g d x_i}{2y_i}, \quad
      d\tilde  u^{(i)}_2 :=  \frac{x_i^{g+1} d x_i}{2y_i}, \quad \cdots,
 \quad
     d\tilde  u^{(i)}_g :=\frac{x_i^{2g-1} d x_i}{2 y_i},
      \tag 2-7
$$
and $d \bold r^{(i)} := (d r_1^{(i)}, d r_2^{(i)}, \cdots, d r_g^{(i)})$,
$$
     (d \bold r^{(i)}):=\Lambda \pmatrix d \bold u^{(i)}
           \\ d \tilde{\bold u}^{(i)}
 \endpmatrix,
     \tag 2-8
$$
where $\Lambda$ is $2g \times g$ matrix defined by
$$
\split
        \Lambda& =
\left(\matrix 0 & \lambda_3 & 2 \lambda_4 &
            3 \lambda_5 & \cdots &
          (g-1)\lambda_{g+1}& g \lambda_{g+2
          }& (g+1)\lambda_{g+3}\\
        \  & 0 &  \lambda_5 & 2\lambda_6 & \cdots &
            (g-2)\lambda_{g+2}& (g-1) \lambda_{g+3}&  g
            \lambda_{g+4}\\
    \ & \ &  0         &  \lambda_7 & \cdots &
            (g-3)\lambda_{g+3}& (g-2) \lambda_{g+4}
               & (g-1)\lambda_{g+5}\\
   \ & \ &  \        & \ & \ddots &
            \vdots & \vdots& \vdots\\
 \ &  & \text{\Large 0}        &  \ & \ &
            \lambda_{2g-2} &2 \lambda_{2g-1}& 3\lambda_{2g+1}\\
 \ &  & \        &  \ & \ &
          0  & \lambda_{2g+1}& 0  \endmatrix\right.\\
&\qquad \qquad
\left.\matrix \cdots & (2g-3)\lambda_{2g-1} & (2g-2)\lambda_{2g}
             & (2g-1) \lambda_{2g+1}\\
          \cdots & (2g-4)\lambda_{2g} & (2g-3)\lambda_{2g+1}
             & 0                    \\
          \cdots & (2g-5)\lambda_{2g+1} &        0
             &                     \\
\cdots & 0 &
             &                  \ \\
     \  & \ &
             &                     \\
\ & \ &  \text{\Large 0}
             &                  \ \\
\ & \ &  \
             &                  \ \endmatrix\right).
\endsplit \tag 2-9
$$

\item The complete hyperelliptic integral matrices
as the second kind are defined by,
$$    \pmb{\eta}':=\left[\left(
         \int_{\alpha_{j}}d r^{(a)}_{i}\right)_{ij}\right],
\quad
      \pmb{\eta}'':=\left[\left(
        \int_{\beta_{j}}d r^{(a)}_{i}\right)_{ij}\right],
 \quad
    \pmb{\omega}:=\left[\matrix \pmb{\omega}' \\ \pmb{\omega}''
     \endmatrix\right].
  \tag 2-5
$$
\item
By defining the Abel map for $g$-th symmetric product
of the curve $C_g$,
$$
\bold u\equiv(u_1,\cdots,u_g)
:\roman{Sym}^g( C_g) \longrightarrow \Bbb C^g,
$$ $$
      \left( u_k((x_1,y_1),\cdots,(x_g,y_g)):= \sum_{i=1}^g
       \int_\infty^{(x_i,y_i)} d u^{(i)}_k \right),
      \tag 2-10
$$
the Jacobi variety  $\Cal J_g$
are defined as complex torus,
$$
   {\Cal J_g} := \Bbb C^g /{ \pmb{\Lambda}} .
     \tag 2-11
$$
Here  ${ \pmb{\Lambda}}$  is a $g$-dimensional
lattice generated by ${\pmb{\omega}}$.

\endroster
\endproclaim

\tvskip

\vskip 0.5 cm
\proclaim {Definition 2-4  }

\it
\roster
The coordinate in $\Bbb C^g$ for
 points $(x_i,y_i)_{i=1,\cdots,g}$
of the curve $y^2 = f(x)$ is given by,
$$
  u_j :=\sum_{i=1}^g\int^{(x_i,y_i)}_\infty d u_j^{(i)} .
    \tag 2-12
$$

\item The hyperelliptic $\sigma$ functions,
which is a holomorphic
function over $u\in \Bbb C^g$, is defined by
[Ba2, p.336, p.350, Kl1, BEL2],
$$ \sigma(u)=\sigma(u;C_g):
  =\ \gamma\roman{exp}(-\dfrac{1}{2}\ ^t\ u
  \pmb{\eta}'{\pmb{\omega}'}^{-1}u)
  \vartheta\negthinspace
  \left[\matrix \delta'' \\ \delta' \endmatrix\right]
  ({\pmb{\omega}'}^{-1}u ;\pmb \tau),
     \tag 2-13
$$
where $\gamma$ is a certain constant factor,
$$\vartheta\negthinspace\left[\matrix a \\ b \endmatrix\right]
     (z; \pmb \tau)
    :=\sum_{n \in \Bbb Z^g} \exp \left[2\pi \sqrt{-1}\left\{
    \dfrac 12 \ ^t\negthinspace (n+a)\pmb \tau(n+a)
    + \ ^t\negthinspace (n+a)(z+b)\right\}\right],
     \tag 2-14
$$
for $g$-dimensional vectors $a$ and $b$,
and
$$
 \delta' :=\ ^t\left[\matrix \dfrac {g}{2} & \dfrac{g-1}{2}
       & \cdots
      & \dfrac {1}{2}\endmatrix\right],
   \quad \delta'':=\ ^t\left[\matrix \dfrac{1}{2} & \cdots
& \dfrac{1}{2}
   \endmatrix\right].
     \tag 2-15
$$

\item
Hyperelliptic
$\wp$-function is defined by [Ba1, Ba2 p.370, BEL2],
$$   \wp_{i j}(u):=-\dfrac{\partial^2}{\partial
   u_i\partial u_j}
   \log \sigma(u) .
         \tag 2-16
$$

\item Hyperelliptic $\al$-function is defined by
[Ba2 p.340, W],
$$
\al_r(u) = \gamma'\sqrt{F(b_r)} , \tag 2-17
$$
where $\gamma'$ is a certain constant factor,
$$
   \split
        F(x)&:= (x-x_1) \cdots (x-x_g)\\
            &= \gamma_g x^g + \gamma_{g-1} x^{g-1} +\cdots + \gamma_{0},
   \endsplit
          \tag 2-18
$$
$\gamma_g\equiv 1$ and $\gamma_i$'s are elementary symmetric functions
of $x_i$'s.

\endroster

\endproclaim

\proclaim{\fp Proposition 2-5}\it

\roster

\item $\wp_{g i}$ $(i=1,\cdots,g)$ is elementary
symmetric functions of $\{x_1,x_2,\cdots, x_g\}$, i.e., for
 $(x_1,\cdots, x_g) \in \roman{Sym} (C^g)$ [Ba1, BEL2],
$$
        F(x) = x^g-\sum_{i=1}^g \wp_{g i} x^{i-1}.
         \tag 2-19
$$

\item
$U:=(2\wp_{gg}+\lambda_{2g}/6)$
 obeys the Korteweg-de Vries
equations [BEL2 p52-53],
 $$
        4\frac{\partial U}{\partial u_{g-1}}
        +6 U \frac{\partial U}{\partial u_g}
               + \frac{\partial^3 U }{\partial u_g}
         =0.\tag 2-20
$$

\endroster

\endproclaim

\proclaim{\fp Definition 2-6}\it
\roster
\item A polynomial associated with $F(x)$ is defined by
$$
\split
\pi_i(x) &:= \frac{F(x)}{x-x_i}\\
        &=\chi_{i,g-1}x^{g-1} +\chi_{i,g-2} x^{g-2}
            +\cdots+\chi_{i,1}x+\chi_{i,0},\\
\endsplit \tag 2-21
$$
where $\chi_{i,g-1}\equiv1$, $\chi_{i,g-2}= (x_1+\cdots+x_g)-x_i$,
and so on.

\item We will introduce $g\times g$-matrices,
$$
 W := \pmatrix
     \chi_{1,0} & \chi_{1,1} & \cdots & \chi_{1,g-1}  \\
      \chi_{2,0} & \chi_{2,1} & \cdots & \chi_{2,g-1}  \\
   \vdots & \vdots & \ddots & \vdots  \\
    \chi_{g,0} & \chi_{g,1} & \cdots & \chi_{g,g-1}
     \endpmatrix,\quad
        \Cal Y = \pmatrix
     y_1 & \ & \ & \  \\
      \ & y_2& \ & \   \\
      \ & \ & \ddots   & \   \\
      \ & \ & \ & y_g  \endpmatrix,
$$
$$
        \Cal F' = \pmatrix
     F'(x_1) & \ & \ & \  \\
      \ & F'(x_2)& \ & \   \\
      \ & \ & \ddots   & \   \\
      \ & \ & \ & F'(x_{g})  \endpmatrix,\quad
\tag 2-22
$$
where $F'(x):=d F(x)/d x$.

\endroster
\endproclaim

\proclaim{\fp Lemma 2-7}\it

\roster

\item The inverse matrix of $W$ is given by $W^{-1}={\Cal F}^{-1} V$,
where $V$ is Vandermond matrix,
$$
        V= \pmatrix 1 & 1 & \cdots & 1 \\
                   x_1 & x_2 & \cdots & x_g \\
                   x_1^2 & x_2^2 & \cdots & x_g^2 \\
                    \vdots& \vdots &       & \vdots \\
                   x_1^{g-1} & x_2^{g-1} & \cdots & x_g^{g-1}
                 \endpmatrix. \tag 2-23
$$

\item By letting $\partial_{u_i}:=\partial/\partial{u_i}$ and
$\partial_{x_i}:=\partial/\partial{x_i}$, we obtain
$$
        \pmatrix \partial_{u_1}\\
                 \partial_{u_2}\\
                 \vdots\\
                 \partial_{u_g}
         \endpmatrix
   =2 \Cal Y \Cal F^{\prime -1}\cdot {}^tW
        \pmatrix \partial_{x_1}\\
                 \partial_{x_2}\\
                 \vdots\\
                 \partial_{x_g}
         \endpmatrix. \tag 2-24
$$
\endroster
\endproclaim

\demo{Proof}
(1) is obvious by direct substitution and uniqueness of
an inverse matrix. We should pay attentions on the
fixed parameters in the partial differential in (2).
$\dfrac{\partial}{\partial u_i}$ means
$\left( \dfrac{\partial}{\partial u_i}\right)_{
     u_1,\cdots,u_{i-1},u_{i+1},\cdots, u_g} $ where indices are fixed
parameters. Due to
$$
        d x_i = \sum_{j=1}^g
 \left(\frac{\partial x_i }{\partial u_j}\right)_{
  u_1,\cdots,u_{j-1},u_{j+1},\cdots, u_g} d u_j,
\tag 2-25
$$
we have
$$
     \left( \frac{\partial}{\partial u_i}\right)_{
     u_1,\cdots,u_{i-1},u_{i+1},\cdots, u_g}
          =\sum_{j=1}^g
\left(\frac{\partial x_j }{\partial u_i}\right)_{
  u_1,\cdots,u_{i-1},u_{i+1},\cdots, u_g}
\frac{\partial }{\partial x_i}. \tag 2-26
$$
Comparing (2-26) with the definition (2-12),
(2) is proved.\qed \enddemo

\tvskip
\centerline{\twobf \S 3.
Hyperelliptic Solutions of Modified Korteweg-de Vries
Equation }\tvskip

In this section, we will give hyperelliptic solutions of the MKdV
equation and  Miura transformation in terms of
theory of hyperelliptic function developed in nineteenth century.

First we will define the modified Korteweg-de Vries (MKdV) equation.

\proclaim{\fp Definition 3-1}\it

The MKdV equation is given by the form,
$$
        \frac{\partial}{\partial t} q
   + 6 q^2 \frac{\partial}{\partial s} q
  + \frac{\partial^3}{\partial s^3} q
         =0, \tag 3-1
$$
where $t$ and $s$ are real or complex numbers.

\endproclaim

Then we will give our main theorem of this article as follows.

\proclaim{\fp Theorem 3-2}\it

\roster

\item By letting
$$\mu^{(r)} = \dfrac{1}{2} \frac{\partial}{\partial u_g} \phi^{(r)},
 \quad
   \phi^{(r)}(u) :=\frac{1}{\sqrt{-1}} \log F(b_r),
        \tag 3-2
$$
$\mu^{(r)}$ obeys the modified KdV equation,
$$
  \left(\frac{\partial}{\partial u_{g-1} }-(\lambda_{2g}+b_r)
          \frac{\partial}{\partial u_g} \right)\mu^{(r)}
           -6 {\mu^{(r)}}^2\frac{\partial}{\partial u_g}  \mu^{(r)}
 +\frac{\partial^3}{\partial u_g^3} \mu^{(r)}=0.
      \tag 3-3
$$

\item We have the Miura relation,
$$
        2\wp_{gg}+\lambda_{2g} + b_r = {\mu^{(r)}}^2
               +\sqrt{-1}\frac{\partial}{\partial u_g}  \mu^{(r)}.
          \tag 3-4
$$

\endroster

\endproclaim

Before proving the theorem, we will remark the meanings of our
theorem.

\proclaim{\fp Remark 3-3}\rm

\roster

\item For arbitrary hyperelliptic curves,
we can construct $\mu^{(r)}$ in (3-2) which obeys the MKdV equation
in terms of the hyperelliptic $\al$-function, $\al_r = \gamma'\sqrt{F(b_r)}$.
It means that we present all finite type solutions of the
MKdV equation which are expressed by meromorphic functions
over hyperelliptic curves up to the symmetries of Proposition 2-2.

\item Using the Miura transformation, the theorem means
that we have another proof of Proposition 2-5 (2)
on the hyperelliptic solutions of the KdV equation.

\item There is a map from $\roman{Sym}^g(C_g)$ to $\CC P^1$
$$\CD
        \{(x_1,x_2,\cdots,x_g)\} @>{\bold u}>>  \Cal J   \\
                @.    @VV{\wp_{gg}}V \\
                @.  \{x_1+\cdots +x_g\}   ,
\endCD \tag 3-5$$
and this characterizes the $\wp_{gg}$ function [BBEIM].
It is noted that we can regard this map as a global affine map
between the affine spaces,
$\roman{Sym}^g(\CC)$ and $\CC$.
The global  translation and dilation
$x_i \to a x_i + b$ for all $i$  ($a, b \in \CC$)
makes  $\wp_{gg} \to  a \wp_{gg} +g b$
but the curve $C_g$ does not change due to Proposition 2-2.
This translation freedom should be one of the keys of the
inverse scattering method [GGKM].
Further Miura transformation connects between
objects which are invariant and variant for translation respectively.
In fact for the translation, Weierstrass al functions are invariant.
It reminds us of stabilizer in homogeneous space.

Further we remark that in the property of infinite point,
$\infty + a = \infty$, the invariance can be translated to
an analysis at infinite point. With the properties of
degenerate curve $y^2 = P(x)^2 x$ as mentioned in [Ma4],
Miura transformation can be interpreted as
B\"acklund transformation
in a certain case.

\item From the point of view of study of symmetric functions,
$F(x)$ is a generator of the elementary symmetric functions
whereas due to (3-9) the definition of $\mu^{(r)}$ in (3-2)
 is the resemble to the generator of the power sum symmetric
function [Mac]
$$
           \frac{ d}{d x} \log F(x) = \frac{1}{x}
           \sum_{j=0}^\infty [\sum_{i=1}^n x_i^j]
            \frac{1}{x^j}. \tag 3-6
$$

\item Due to the Miura transformation, the differential
$\partial \mu^{(r)}/\partial u_g$
can be expressed by polynomial of $\wp_{gg}$ and $\mu^{(r)}$'s.
In other words, it connects categories of
analysis and algebra. In fact, the Miura
transformation can be translated to
a language in theory of
a geometrical Dirac equation of a curve in a complex
plane, or Frenet-Serret relation [Ma4].
There the Miura transformation is interpreted as an integrability
condition in the $\Cal D$-module ([Bj] p.12-13, [Ma4]).

\item
In [EEK], Eilbeck, Enolskii and Kostov studied the
vector nonlinear Schr\"osinger equations and obtained a
formula, ((3-8) in [EEK]) which is essentially the same as (3-3),
even though they did not mention a relation between
their formula  and the Miura transformation.

\endroster

\endproclaim

Here we will mention our plan to prove the theorem.
First we will prove the Miura relation (3-4).
As the first step in its proof, we will
translate the operation of $\partial/\partial u_g$ into
that of $\partial / \partial x_i$ and compute (3-2).
Secondly we will evaluate the obtained terms by using the symmetries of
summation and residual computation over the curve.
Then we obtain the Miura relation (3-4).
Next, we will investigate the the right hand side in
an integration version of (3-3). If we show that
$\phi^{(r)}$ in (3-2) obeys
$$
4\left(\frac{\partial}{\partial u_{g-1} }-(\lambda_{2g}+b_r)
          \frac{\partial}{\partial u_g} \right)\phi^{(r)}
+ \frac{1}{2}
\left(\frac{\partial}{\partial u_g} \phi^{(r)}\right)^3
     + \frac{\partial^3}{\partial u_g^3} \phi^{(r)}=0,
           \tag 3-7
$$
we have a solution of (3-3) by differentiating (3-7) in $u_g$ again.
In the derivation, we will use the Miura relation and a
correspondences between $\partial/\partial u_{g-1}$ and
$\partial / \partial x_i$'s.
We will obtain (3-7) at (3-31).

Now let us prove the theorem.
From (2-25), we will express $u$'s by the affine coordinate
$x_i$'s,
$$
        \frac{\partial}{\partial u_g }=
         \sum_{i=1}^g \frac{2y_i}{F'(x_i)} \frac{\partial}{\partial x_i},
           \quad
        \frac{\partial}{\partial u_{g-1} }=
         \sum_{i=1}^g \frac{2y_i\chi_{i,g-1}}{F'(x_i)}
              \frac{\partial}{\partial x_i}.
       \tag 3-8
$$
Hence we have
$$
\mu^{(r)}\equiv \frac{1}{2\sqrt{-1}}
        \frac{\partial}{\partial u_g }\log F(b_r)
       =\frac{1}{\sqrt{-1}}
\sum_{i=1}^g\frac{y_i}{F'(x_i) (x_i-b_r)}
  =\frac{1}{\sqrt{-1}}
\sum_{j=1}^\infty\sum_{i=1}^g\frac{y_i}{F'(x_i) b_r }
\frac{ x_i ^j}{b_r^j},\tag 3-9
$$ $$
        \frac{\partial}{\partial u_{g-1} }\log F(b_r)
         =\sum_{i=1}^g\frac{2y_i\chi_{i,g-1}}{F'(x_i)( x_i-b_r)}
          . \tag 3-10
$$
Roughly speaking, we wish to compute
$$
\frac{1}{2}\frac{\partial^2}{\partial u_g^2 }\log F(b_r)
       =
         \sum_{j=1, i=1}^g \frac{y_j}{F'(x_j)} \frac{\partial}{\partial x_j}
         \frac{2y_i}{F'(x_i) (x_i-b_r)},
          \tag 3-11
$$
and express it by lower derivative.

Here we will summary the derivatives of $F(x)$, which are
shown by direct computations.

\proclaim{\fp Lemma 3-4}\it

\roster

\item
$$
        \frac{\partial}{\partial x_i} F(x) =- \frac{F(x)}{x-x_j}
           .      \tag 3-12
$$

\item
$$
        \left[\frac{\partial}{\partial x} F(x)\right]_{x=x_i}
           =\prod_{j\neq i} (x_i-x_j)
           .      \tag 3-13
$$

\item
$$
        \frac{\partial}{\partial x_k} F'(x_k)
         =\frac{1}{2}
 \left[\frac{\partial^2}{\partial x^2} F(x)\right]_{x=x_k}
         . \tag 3-14
$$

\endroster
\endproclaim

In order to evaluate (3-11), we will first show the next lemma.

\proclaim{\fp Lemma 3-5}\it

\roster

\item
$$
\sum_{k=1}^g \frac{1}{F'(x_k)}
             \left[\frac{\partial}{\partial x}\left(
       \frac{f(x)}{(x - b_r) F'(x)} \right) \right]_{x = x_k}
       = \lambda_{2g} + b_r + 2 \wp_{gg}
         . \tag 3-15
$$

\item
$$
\left( \sum_k \frac{y_k}{(x-x_k) F'(x_k)} \right)^2
  =\sum_{k,l, k\neq l}
     \frac{2y_k y_l}{(x-x_k) (x_k - x_l)F'(x_k)F'(x_l)}
  +\sum_k \frac{y_k^2}{(x-x_k)^2 F'(x_k)^2}
         . \tag 3-16
$$
\endroster

\endproclaim

\demo{Proof}:
In order to prove  (1), we will consider an integral
over a contour in $C_g$ and step by four processes as follows.
\roster

\item Let $\partial C_g^o$ be a boundary of a polygon representation
$C_g^o$ of the curve $C_g$,
$$
  \oint_{\partial C_g^o} \frac{f(x)}{(x-b_r)F(x)^2} dx =0
 .           \tag 3-17
$$

\item The divisor of the integrand of (1) is
$$
\left(\frac{f(x)}{(x-b_r)F(x)^2} dx\right) =
     \sum_{i=1, b_i \neq b_r}^{2g+1} (b_i,0) -
        2\sum_{i=1}^g (x_i,y_i) -
        2\sum_{i=1}^g (x_i,-y_i) - 3\infty . \tag 3-18
$$

\item Noting that the local parameter $t$ at $\infty$ is $x=1/t^2$,
$$
\res_{\infty}\frac{f(x)}{(x-b_r)F(x)^2} dx
      = -2 (\lambda_{2g} + b_r + 2 \wp_{gg}). \tag 3-19
$$

\item Noting that the local parameter $t$ at $(x_k,\pm y_k)$ is
$t=x-x_k$,
$$
\res_{(x_k,\pm y_k)}\frac{f(x)}{(x-b_r)F(x)^2} dx
      =  \frac{1}{F'(x_k)}
             \left[\frac{\partial}{\partial x}\left(
       \frac{f(x)}{(x - b_r) F'(x) }\right) \right]_{x = x_k}
        . \tag 3-20
$$

\endroster
By arranging them, we obtain (3-15).
On the other hand, (2) can be proved by using a  trick:
 for $i\neq j$,
$$
         \frac{1}{(x_j-x_i) (x_i-b_r)}+ \frac{1}{(x_i-x_j) (x_j-b_r)}=
         \frac{1}{(x_j-b_r) (x_i-b_r)}.
          \tag 3-21
$$
Then we have (3-16).
\qed\enddemo

Let us compute (3-11), which is $\sqrt{-1}\partial \mu^{(r)} /
\partial u_g$,  as follows;
$$
        \frac{1}{2}\frac{\partial^2}{\partial u_g^2 }\log F(b_r)
       =
         2\sum_{i=1}^g \frac{y_i}{F'(x_i)} \frac{\partial}{\partial x_i}
         \frac{y_i}{F'(x_i) (x_i-b_r)}
       + 2 \sum_{i=1,j=1,j\neq i}^g \frac{y_j}{F'(x_j)}
          \frac{\partial}{\partial x_j}
         \frac{y_i}{F'(x_i) (x_i-b_r)}.
          \tag 3-22
$$
The first term in the right hand side of (3-22) becomes
$$
         \sum_{i=1}^g \frac{y_i}{F'(x_i)}\left(
         \frac{1}{y_iF'(x_i) (x_i-b_r)} \frac{d f(x_i)}{d x_i}
         -\frac{y_i}{F'(x_i) (x_i-b_r)} \left[
           \frac{\partial}{\partial x} F(x) \right]_{x=x_i}
       -  \frac{2y_i}{F'(x_i) (x_i-b_r)^2}\right).
          \tag 3-23
$$
Using the lemma 3-4 (1), it can be modified to,
$$
         \sum_{i=1,j=1,j\neq i}^g  \frac{1}{F'(x_k)}
             \left[\frac{\partial}{\partial x}\left(
       \frac{f(x)}{(x - b_r) F'(x) }\right) \right]_{x = x_i}
       -
         \sum_{i=1}^g \frac{f(x_i)}{F'(x_i)^2}\left(
         \frac{1}{ (x_i-b_r)^2}\right).
          \tag 3-24
$$
Using (3-15), the first term in (3-24) is expressed well.
On the other hand, the second term in the right hand side of
(3-22) is computed to
$$
       2 \sum_{i=1,j=1,j\neq i}^g \frac{y_i y_j}{F'(x_j)F'(x_i)}
         \frac{1}{(x_j-x_i) (x_i-b_r)}.
          \tag 3-25
$$
From lemma 3-5 (2), the second term in (3-24) and (3-25) becomes
$$
       \frac{1}{4}  \left(\sum_{i=1}^g \frac{2y_i}{F'(x_i)}
         \frac{1}{(x_i-b_r)}\right)^2\equiv -{\mu^{(r)}}^2.
          \tag 3-26
$$
Hence we have Theorem 3-2 (2).

Next we will prove (3-3) as follows:
From theorem 3-2 (2), $(\gamma_{g-1}\equiv-\wp_{gg}$ due to (2-18) and (2-19))
,
we have
$$
 \frac{\partial}{\partial u_g} \mu^{(r)}=\frac{1}{\sqrt{-1}}
\left(-2\gamma_{g-1}+\lambda_{2g} + b_r - {\mu^{(r)}}^2\right)
               .
          \tag 3-27
$$
We perform the differential
in $u_g$ to both hand side again,
$$
\split
 \frac{\partial^2}{\partial u_g^2} \mu^{(r)}&=
 \frac{1}{\sqrt{-1}}\frac{\partial}{\partial u_g}
         \left(-2\gamma_{g-1}+\lambda_{2g} + b_r - {\mu^{(r)}}^2\right)\\
&=
 \frac{1}{\sqrt{-1}}
         \left(-2\sum_{i=1}^g \frac{2y_i}{F'(x_i)}- 2{\mu^{(r)}}
                       \frac{\partial}{\partial u_g} \mu^{(r)}\right)\\
&=
 \frac{-2}{\sqrt{-1}}
         \left(\sum_{i=1}^g \frac{2y_i}{F'(x_i)}+
\frac{1}{\sqrt{-1}}{\mu^{(r)}}
     \left[-2\gamma_{g-1}+\lambda_{2g} + b_r -
     {\mu^{(r)}}^2\right]\right).\\
\endsplit\tag 3-28
$$
In the second step, we used (3-27) again and
$$
\frac{\partial}{\partial u_g}
       \gamma_{g-1}
=-\sum_{i=1}^g \frac{2y_i}{F'(x_i)}.\tag 3-29
$$
Noting $\chi_{i,g-1}=\gamma_{g-1} -x_i$,
the first and second term in the parenthesis is
$$
\split
\sum_{i=1}^g \frac{2y_i}{F'(x_i)}-
\frac{2}{\sqrt{-1}}\gamma_{g-1}{\mu^{(r)}}&
=\sum_{i=1}^g \frac{2y_i}{F'(x_i)}\frac{x_i-b_r}{x_i-b_r}
+ \gamma_{g-1}
\sum_{i=1}^g \frac{2y_i}{F'(x_i)(x_i-b_r)}\\
&=\frac{\partial}{\partial u_{g-1}}\log F(b_r)
-\frac{2}{\sqrt{-1}} b_r{\mu^{(r)}}.
\endsplit\tag 3-30
$$
Thus we obtain (3-7),
$$
-\frac{1}{4}\frac{\partial^3}{\partial u_g^3} \phi^{(r)}=
        \left(\frac{\partial}{\partial u_{g-1} }-(\lambda_{2g}+b_r)
          \frac{\partial}{\partial u_g} \right) \phi^{(r)}
          + \frac{1}{8}
\left(\frac{\partial}{\partial u_g} \phi^{(r)}\right)^3.
\tag 3-31
$$
We differentiate (3-31) in $u_g$ again and then
obtain (3-3). Hence we completely prove our theorem 3-2.

\vskip 0.5 cm

{\centerline{\bf{\S 4 Discussion}}}

\vskip 0.5 cm
We obtained the hyperelliptic solutions of MKdV equation
in terms of hyperelliptic $\al$ functions without any theta functions,
which is in contrast to the approach of Gesztesy and Holden [GH1]
and others [BBEIM and references therein].
 For an arbitrary
hyperelliptic curve, we can write down its
explicit function form by $\al$ function.
Miura transformation means  a connection between
the $\wp_{gg}$-functions and $\mu^{(r)}$'s in (3-2)
which are differentials of the hyperelliptic al-functions;
this aspect also differs from study on the Miura transformation
of  Gesztesy and Holden [GH2].

According to [Kl2] and [D], Weierstrass found the $\sigma$ function
for an elliptic curve, which he called $al$-function
as a honor of Abel, and his approach  was to investigate
divisor decomposition of a function,
$$
     y = 4 y_1 y_2 y_3, \tag 4-1
$$
for an elliptic curve $y^2 = 4 x^3 + g_3 x + g_4=4(x-e_1)(x-e_2)(x-e_3)$;
$y_i =  \sqrt{x - e_i}$. These $y_i$ differs
from his $\sigma$-function itself but corresponds to
our al-function (2-17). $y_i$
 is expressed by a ratio of two $\sigma$'s.
The function $y_i$'s are generators
of Jacobi sn, cn, dn functions.  Comparison of $y_i$'s with
 $\wp$ gives us fruitful information for the elliptic curve case.
It is useful to identify these $sn$'s with $y_i$'s.

For the case of hyperelliptic curves, Weierstrass called
$y_i$'s themselves $al$-functions and defined them as in (2-17)
with explicit constant factor $\gamma'$ [D, Kl, W].
(The hyperelliptic sigma functions also
are factors of al${}_r$'s, al${}_r(u)
=\sigma(u+B_r)/\sigma(u)$ for a certain $B_r$,
which were discovered by Klein [Kl1].)
In [Ba2], Baker found the KdV hierarchy and KP equation
around 1898 by comparing  $F(x)$ and $\wp$'s.
It is also important that we roughly identified with
$F(b_r)$ and al${}_r$ and recognize that  $\mu^{(r)}$
comes from $F(b_r)$.
Miura transformation is a key of the relations
between $F(x)$ and $\wp_{gg}$
as shown in this article. In fact, the techniques in
the proof of Theorem 3-2 are essentially contained in [Ba2].
In other words we should regard that there existed "Miura
transformation" behind discoveries of
KdV hierarchy and KP equation by Baker, from first [Ba2].

In theory of elliptic functions, we know, by experiences,
that Jacobi elliptic functions rather than $\wp$ function
play important roles in physics
whereas $\wp$ is more essential in number theories and
algebraic geometry.
I stemmed a question why there are such difference
between Jacobi elliptic functions, special ones of al-functions,
and $\wp$ function.
One of answers is that al function is
invariant for the affine transformation as mentioned in
Remark 3-3. On the other hand, $\wp$ is not. Of course
for the standard representation of elliptic curve, {\it i.e.},
$y^2 = 4 x^3 - g_2 x + g_3$,
in order to make the coefficient of $x^2$ vanish,
the translation freedom is constraint. However we can find
$\wp$ even for $y^2 = 4 x^3 + g_1 x^2+ g_2 x + g_3$ and
the nature of such $\wp$ is $x$ itself in the affine space.
Thus  al and $\wp$ live in different categories.
As we show in [Ma1] and [Ma4], al is associated with
the differential geometry and solutions of Dirac equations,
while $\wp$ is connected to
algebraic geometry;
$\wp$ is the affine coordinate of the curve.

Hence we conclude that Miura transformation is a connection
between the worlds of $\wp$ and al's.
I think that the researchers in nineteenth century
might implicitly already recognized these facts.
However as we can, now,
look around both theories of "Abelian functions theories"
in last century and
nineteenth century, I think that we should unify them and
develop a new theory beyond them.

\vskip 0.5 cm

{\centerline{\bf{ Acknowledgment}}}

\vskip 0.5 cm

I'm grateful to  Prof.~Y.~\^Onishi for many suggestions
for this work.
I thank Prof. V. Z. Enolskii for sending
his interesting works and giving helpful comments.

\enddocument